\documentclass[a4paper]{jpconf}
\usepackage{graphicx}
\usepackage{subfig}

\begin{document}
\title{Hard Quasi-real Photo-production of Charged Hadrons at COMPASS energies}

\author{Astrid Morr\'eale \sf{on behalf of the COMPASS collaboration}}

\address{American National Science Foundation and French Alternative Energies and Atomic Energy Commission, CEA Saclay, IRFU/SPhN, Gif sur Yvette, 91191, France}

\ead{astrid.morreale@cern.ch}

\begin{abstract}
The Common Muon Proton Apparatus for Structure and Spectroscopy (COMPASS) at CERN with its use of beams of naturally polarized muons scattered of a polarized deuteron target, provides an environment of hard scattering between quasi-real photons and partons.
Hard hadron quasi-real photo-production with polarized initial states is sensitive to the polarized gluon distribution $\Delta$G through $\gamma$-gluon($g$) direct channels as well as $q$-$g$ resolved processes.  Comparisons of unpolarized differential cross section measurements to next-to-leading order (NLO) pQCD calculations are essential to develop our understanding of proton-proton and lepton-nucleon scattering at varying center of mass energies.  These measurements are important to asses the applicability of NLO pQCD in interpreting polarized processes.  In this talk we will discuss unidentified charged separated hadron production at low $Q^{2}$ (Q$^{2}<0.1~GeV^{2}/c^{2}$) and high transverse momenta (p$_{T}>1.0\,GeV/c$). $<p_{T}^2>$ spectra of charged hadrons at $Q^{2}>1~GeV^{2}/c^{2}$ will also be discussed.
\end{abstract}

\section{Introduction}
Measurements sensitive to the gluon's polarized density function $\Delta G$ can be performed by looking at cross-sections of final state inclusive and semi-inclusive particles.  
While perturbative quantum chromo dynamics (pQCD) is a well established theory of strong interactions, it does not provide an exact analytical calculation. In a pQCD factorized framework, a hadron cross-section can be defined as the convolution of three main ingredients (Eq.~\ref{formula_xsect}) from which only one is fully calculable:  
\begin{itemize}
\item The parton distribution functions $f_{a}(x_{a},\mu_F)$,  $f_{\gamma}(x_{\gamma},\mu_F)$ and $\Delta f_{a}(x_{a},\mu_F)$ at a factorization scale $\mu_F$ and momentum fraction $x_{a}$ carried from parton $a$ (or $\gamma$.)
\item The fragmentation functions (FF) $D_{h^{\pm}}(z_{b},\mu_F)$ where $z_{b}$ is the relative energy of hadron $h$ proceeding from parton $b$.\item The hard scattering cross-section of partons $d\hat{\sigma^b_{\gamma a}}(d\Delta\hat{\sigma^b_{\gamma a}})$ which is the calculable part in pQCD.  

\end{itemize}
In practice, the extraction of polarized density information relies on measurements of asymmetries of final state particles in bins of measured transverse momentum p$_{T}$. Asymmetries are the ratio of the polarized to unpolarized cross-sections (Eq.~\ref{formula_all}). These measurements offer an elegant way of accessing parton information by counting observed particle yields in different helicity states of the interacting beam(s) and target ($++, --$\, versus\, $+-, -+$.) These are normalized by the polarization in the beam and the target ($P_{B}, P_{T}$) times the dilution factor $F$.

\begin{eqnarray}
d\sigma &=&\sum_{a, b= {\rm q}(\bar{\rm q}), {\rm g}}\,
\int dx_\gamma
\int dx_a
\int dz_b \,\,
 f_\gamma (x_\gamma,\mu_F) \, f_a (x_a,\mu_F)
D_b^{h}(z_b,\mu_F')\times d\hat{\sigma^b_{\gamma a}}  
\label{formula_xsect}\end{eqnarray}

\newcommand{\sz}{\hspace*{-6pt}}
\begin{eqnarray}
 A_{LL} &\sz=\sz&  \frac{\displaystyle\sum_{a, b={\rm q}(\bar{\rm q}), {\rm g}} 
\Delta f_{\gamma }\otimes \Delta f_{a} \otimes \Delta \hat{\sigma}\otimes D_{h/b}} 
{\displaystyle\sum_{a, b={\rm q}, \bar{\rm q}, {\rm g}} f_{\gamma} 
\otimes f_{a} \otimes \hat{\sigma} 
\otimes D_{h/b}} = \frac{\sigma_{++} -\sigma_{+-}}{\sigma_{++}+\sigma_{+-}} \,\, 
%\nonumber
=\,\, \frac{1}{FP_{B}P_{T}} \frac{N^{++}-N^{+-}}{N^{++}+N^{+-}} 
\qquad\qquad  
\label{formula_all}
\end{eqnarray}

 Unpolarized particle production measurements form the denominator of the asymmetries definition. These unpolarized measurements add new results at different $\sqrt{s}$ than those available~\cite{soffer}. In addition to verifying applicable theoretical calculations, cross-sections can help constrain the FF's~\cite{dss} which introduce an uncertainty into these models. The aim of the current work is to contribute with measurements of semi-inclusive deep inelastic scattering (SIDIS) cross-sections at $\sqrt{s}=\,17 \,GeV$. Unpolarized SIDIS probes the number density of partons  with a fraction $x$ of the momentum of the parent nucleon. The detection of the outgoing lepton in a SIDIS measurement allows the access, among other kinematic variables, to the $Q^{2}$ and $y$ of the events and not just the $p_{T}$ of the final state hadrons.  

\section{Hadron production at $\sqrt{s}=17\,GeV$}
\begin{figure}
\centering
\subfloat[$\gamma_{dir}$ and $\gamma_{res}$ hadron production contributions of different partonic channels ($a$, $b$) as a function of $p_{T}$. Figure is taken from~\cite{jager2}]{\label{werner}\includegraphics[width=0.40\textwidth]{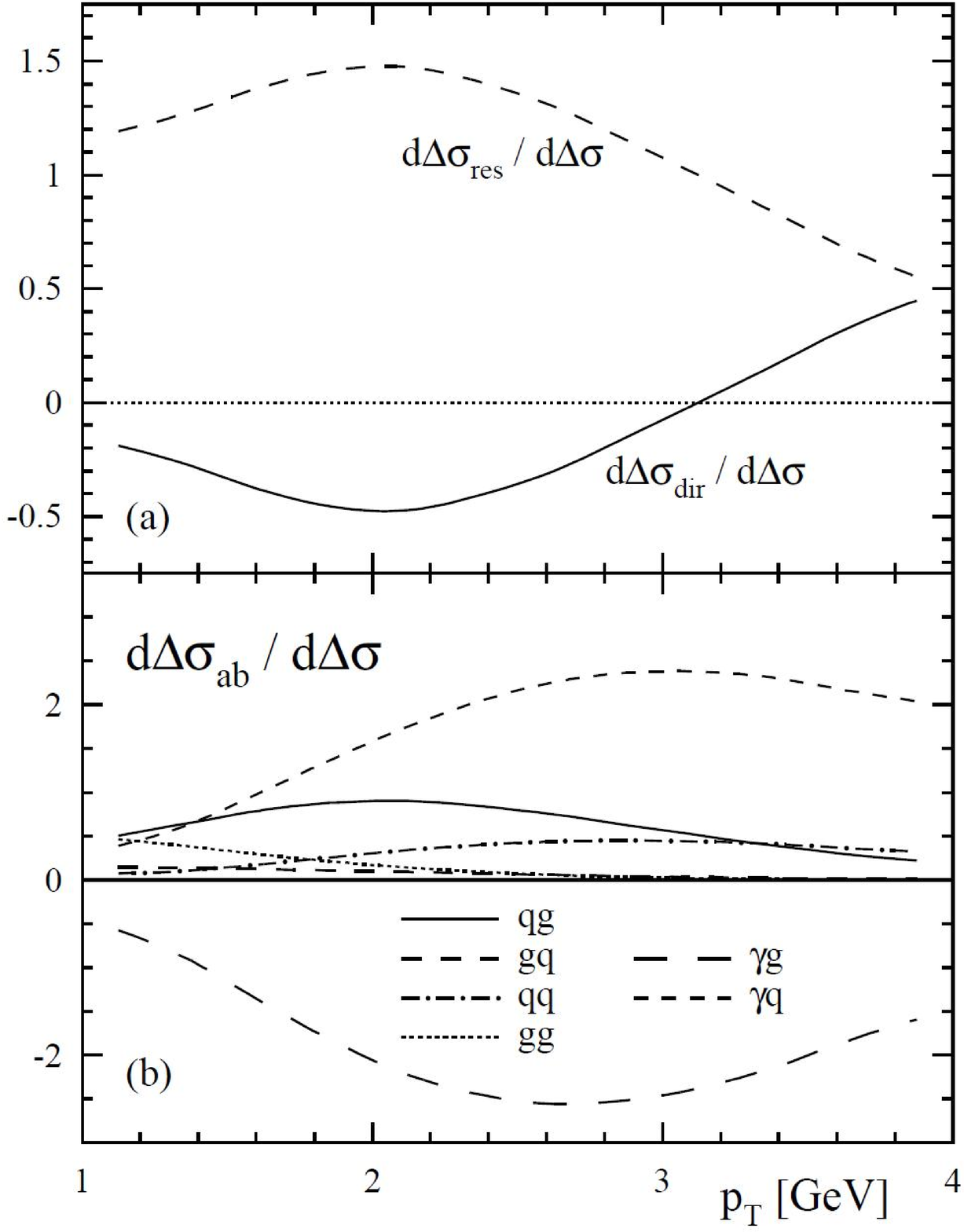}}\,\,\,\,\,\,
\subfloat[Production cross-section as a function of $p_{T}$ for different center of mass energies and compared to pQCD calculations. Figure is taken from~\cite{soffer}.]{\label{jacques}\includegraphics[width=0.40\textwidth]{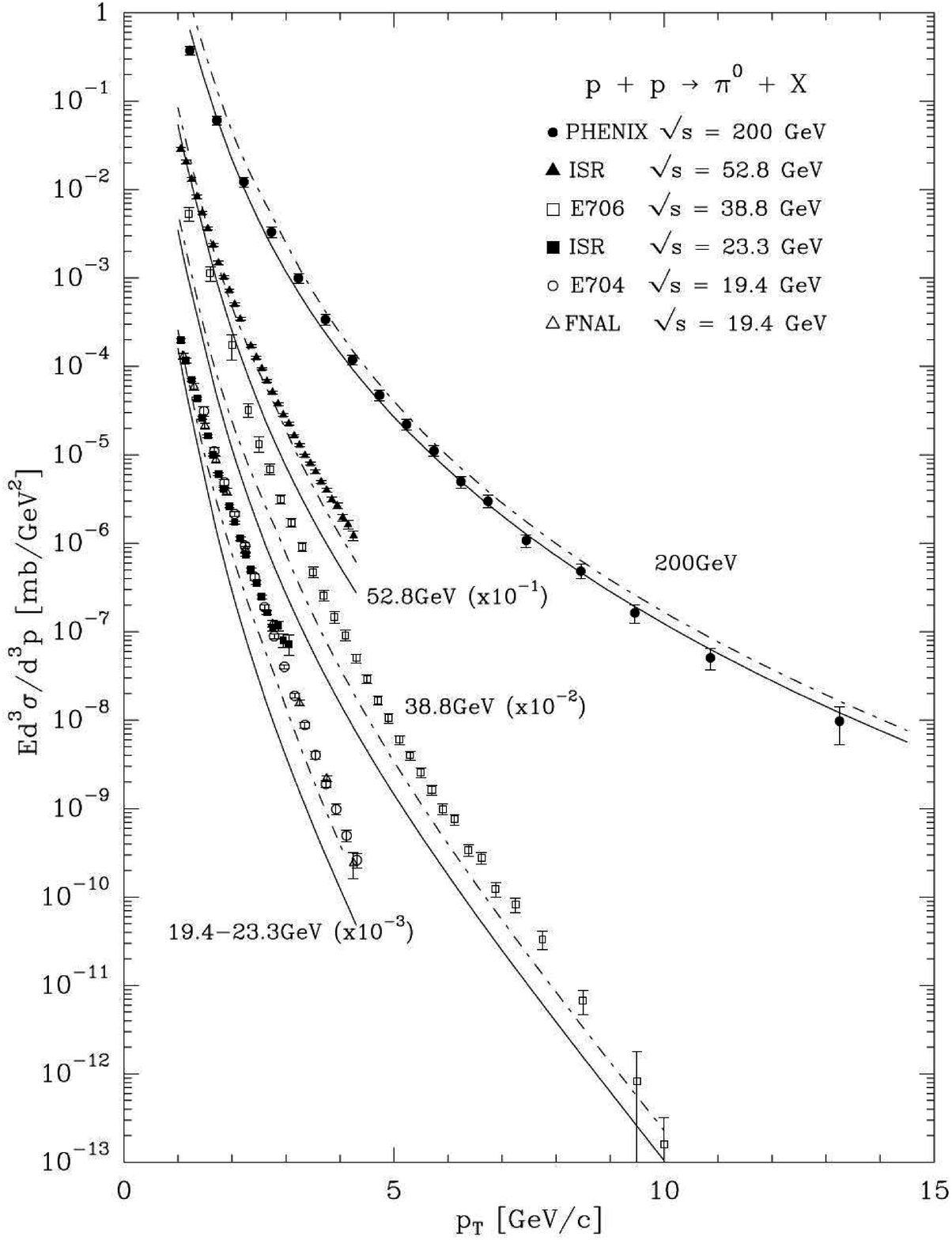}}
 \vspace{-10pt}
\caption{Partonic contributions to charged hadron production at $\sqrt{s}=17\,GeV$ (a). Measurements of $\pi$ production at different values of $\sqrt{s}$ (b).}
\end{figure}

The reaction of interest for the current measurement is $\mu^{+} d\longrightarrow \mu^{+'}h^{\pm}X$. Final state hadrons are detected at low $Q^{2}$ (below $0.1~GeV^{2}/c^{2}$ ) and high $p_{T}$ (above 1~GeV/c). This latter ensures a large momentum transfer in the reaction~\cite{jager1}. 
Production of these final state hadrons proceed from direct photon-parton $\gamma_{dir}-(q,\bar q,g)$ and resolved photon-parton $\gamma_{res}-(q,\bar q, g)$  initiated sub-processes in $\mu-{\rm deuteron}$ center of mass (cm) collisions (Fig.~\ref{werner}.) The  $\gamma_{dir}-(q,\bar q,g)$ NLO contributions are those for which the hadrons are produced immediately after the virtual photon interacts directly with the target partons. The  $\gamma_{res}-(q,\bar q, g)$ processes are those where the photon fluctuates into partons before interacting with the target. These latter processes are akin to hadroproduction  from proton-proton collisions such as those at RHIC or at the LHC, where the partons from the incoming beam interact with the target(beam) partons. Previous comparisons of pQCD calculations for hadron production at different experiments and  $\sqrt{s}$ ranges have been done in~\cite{soffer} (Fig.~\ref{jacques}.) The results of these comparisons showed that the pQCD calculations could not correctly describe hadron production proceeding from proton-proton data throughout all the different $\sqrt{s}$. Only collisions at higher energies seemed to be properly described by the calculations. The failure of the calculations to describe hadron production at lower center of mass energies has opened the question of pQCD applicability at these energy regimes. Other issues which have been raised point out that perhaps the experiments at different center of mass energies do not correspond to the same physics phenomena, in particular at forward rapidity values. It is our present aim to compare the COMPASS measurements to NLO calculations which take into account full COMPASS kinematics. These partonic contributions and predictions at COMPASS energies have been previously calculated in~\cite{jager1} and~\cite{jager2}. 

%\begin{figure}
%\begin{minipage}{16pc}
%\includegraphics[width=15.5pc]{soffer1.eps}
%\caption{\label{jacques}Production cross-section as a function of $p_{T}$ for different center of mass energies and compared to pQCD calculations. Figure is taken from~\cite{soffer}.}
%\end{minipage} 
%\end{wrapfigure}

\section{The COMPASS detector at CERN}
The COMPASS experiment is a fixed target experiment at CERN which uses secondary and tertiary beams from the Super Proton Synchrotron (SPS). The muon beam used for the scattering comes from the decay of  $\pi^{+}$ which proceed from the 
scattering of a primary proton beam on a Beryllium target. The proton beam is extracted from the CERN's SPS at a cycle of approximately 17 seconds. The resulting muon beam from each cycle is then collided on a two cell $^6LiD$ target. The two stage forward spectrometer tracks particles from these interactions with a rate capability of a few MHz/channel~\cite{compass} and a spacial precision better than $100~\mu m$. Particles can be detected at very small angles ($\sim1~$mrad) and large angles ($\sim$ 120~mrad for the current data set under study.)  These scattering angles correspond to a primarily positive pseudo-rapidity coverage ($\eta_{cm}$) that can reach up to the forward values of $\sim2$. For the low $Q^{2}$ measurements of interest the ladder and inner triggers~\cite{compass} are used.
\begin{figure}

\centering
\begin{minipage}{35pc}
\includegraphics[width=35pc]{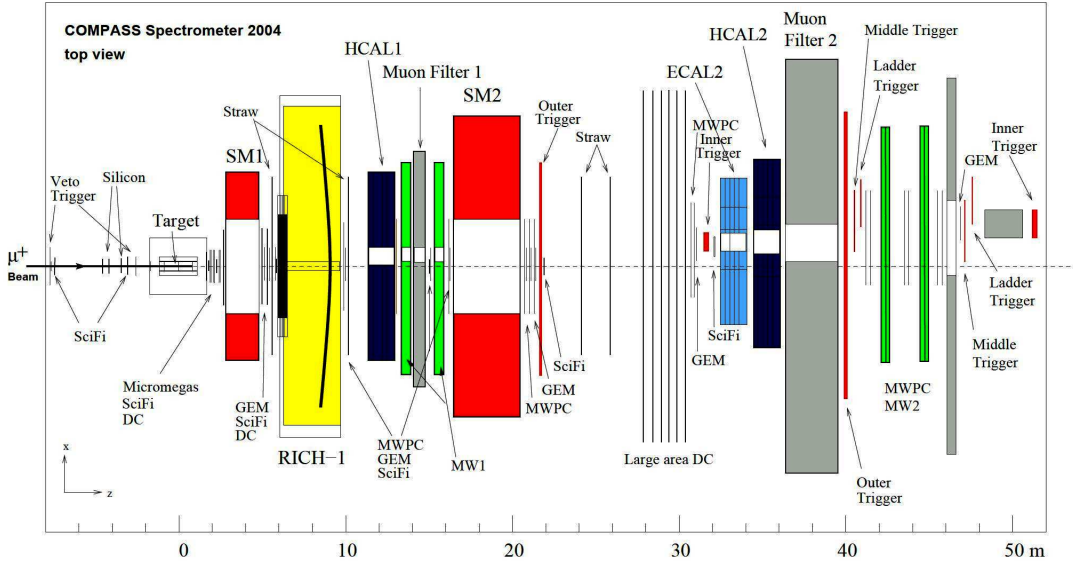}
\vspace{-20pt}
\caption{\label{compass}Artistic top view of the COMPASS 2004 $\mu-d$ spectrometer setup. Figure from~\cite{compass}.}
\vspace{-10pt}
\end{minipage} 
\end{figure}
\section{Measuring cross-sections at COMPASS}
To measure production cross-sections,  the number of all final state charged hadrons in the data-set under interest are normalized by detector acceptances and efficiencies as well as the total integrated luminosity $\int Ldt$. The experimental definition which can be derived from Eq.~\ref{formula_xsect} is as follows:

\begin{eqnarray}
E\frac{d^{3}\sigma}{d\vec{p}^{3}} 
&=& \frac{d^{3}\sigma}{p_{T}d\eta dp_{T}d\phi} = \frac{1}{2\pi p_{T}} \frac{d^{2}\sigma}{dp_{T}d\eta}\,= \frac{1}{2\pi p_{T}} \frac{N_{h^{\pm}}(p_{T})}{\int Ldt\,\epsilon_{Acc}\Delta p_{T} \Delta \eta}\label{formula_xsect2}
\end{eqnarray}

Where:\\

\begin{itemize}\label{itemsxsec}
\item \textit{$\vec{p}^3$} is the 3 momentum of the particle.
\item $\eta$ is the pseudo-rapidity, described as $\eta =-ln[tan(\frac{\theta}{2})]$, where $\theta$ is the angle between the
 particle momentum $\vec{p}$ and the virtual photon axis. $\eta$  can be translated to the center of mass frame to:\\$\eta_{cm}=-ln(tan(asin(p_{T}/P)/2))-0.5ln(2P_{beam}/M_{proton})$.%\\\\As only a coherent scattering off an individual nucleon is considered, only the proton's mass is needed in the above formula. 
 
\item p$_{T}$ is the transverse momentum of the particle, defined with respect to the virtual photon.
\item $\Delta$p$_{T}$ is the bin width.
\item $\phi$ is the azimuthal angle ($2\pi$ in the case of COMPASS.)
\item N$_{h^{\pm}}$ is the total number of reconstructed final state hadrons (charge separated) in a p$_{T}$ bin.
\item $\epsilon_{Acc}$ accounts for the geometrical acceptance, momentum smearing, as well as the efficiency of the reconstruction algorithm. 
\item $\int Ldt$ is the integrated luminosity. 
\end{itemize}
\subsection{Simulation}
\begin{figure}[h]
\centering
\begin{minipage}{28pc}
\includegraphics[width=14pc]{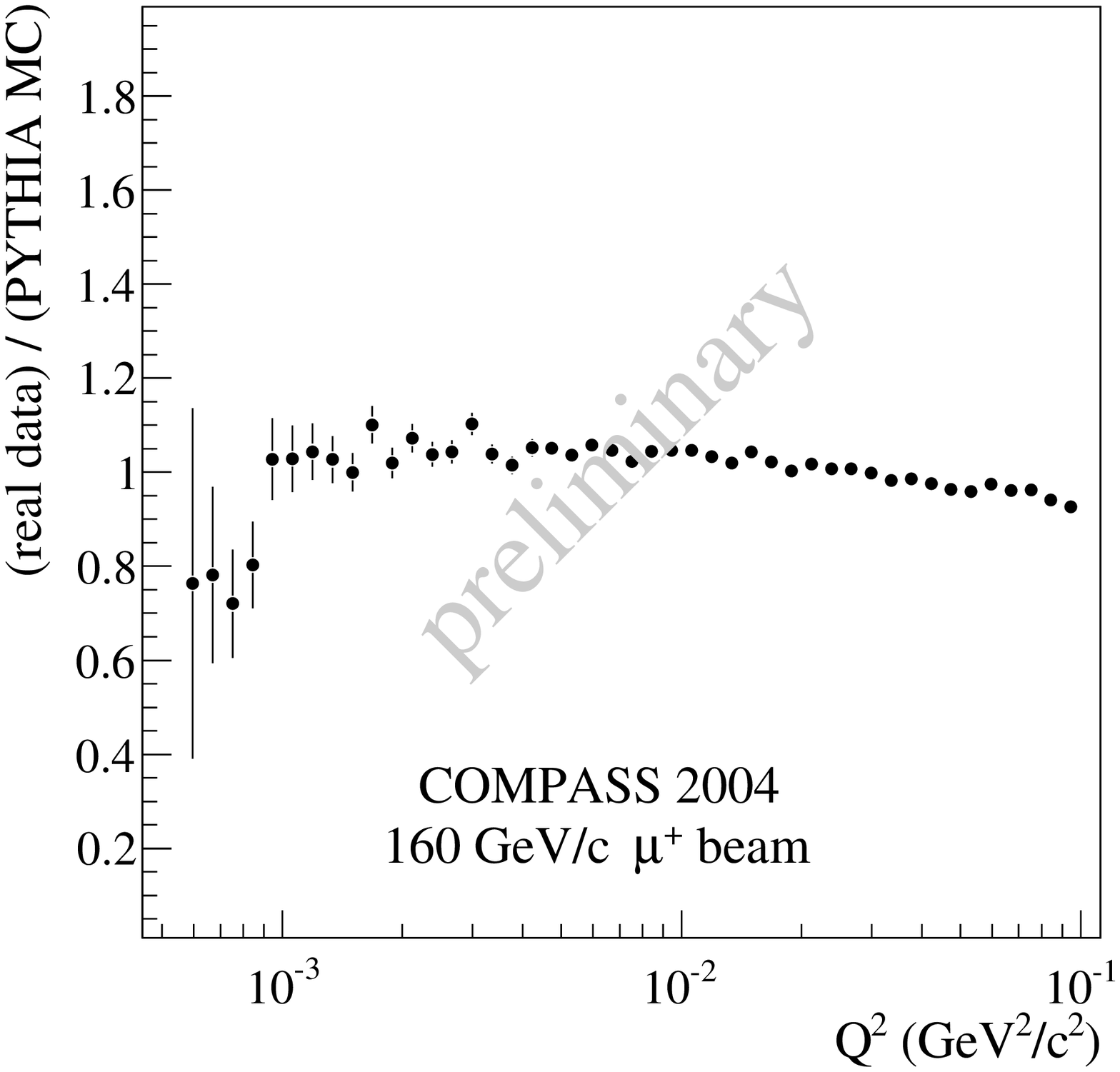}\includegraphics[width=14pc]{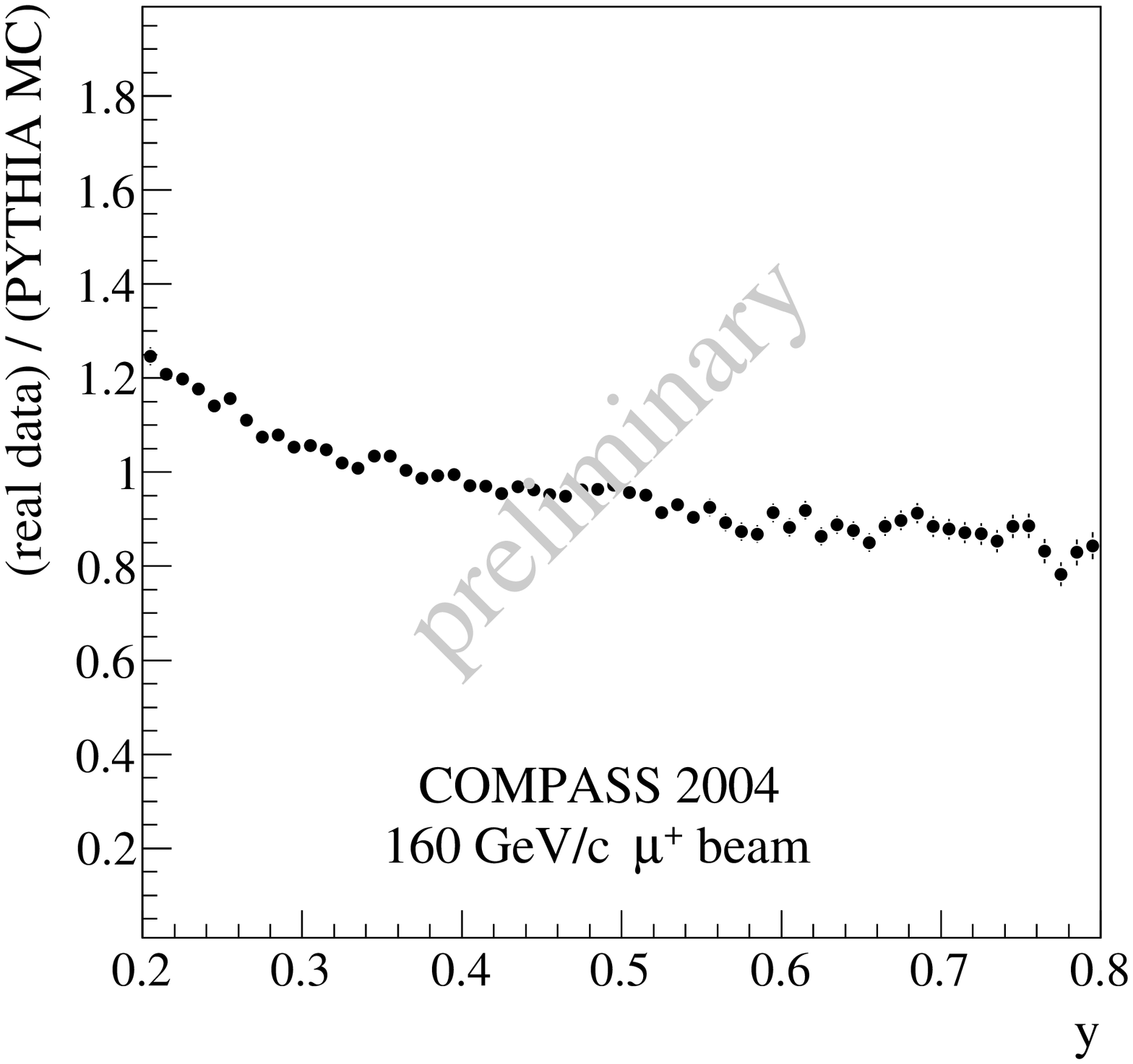}
\includegraphics[width=14pc]{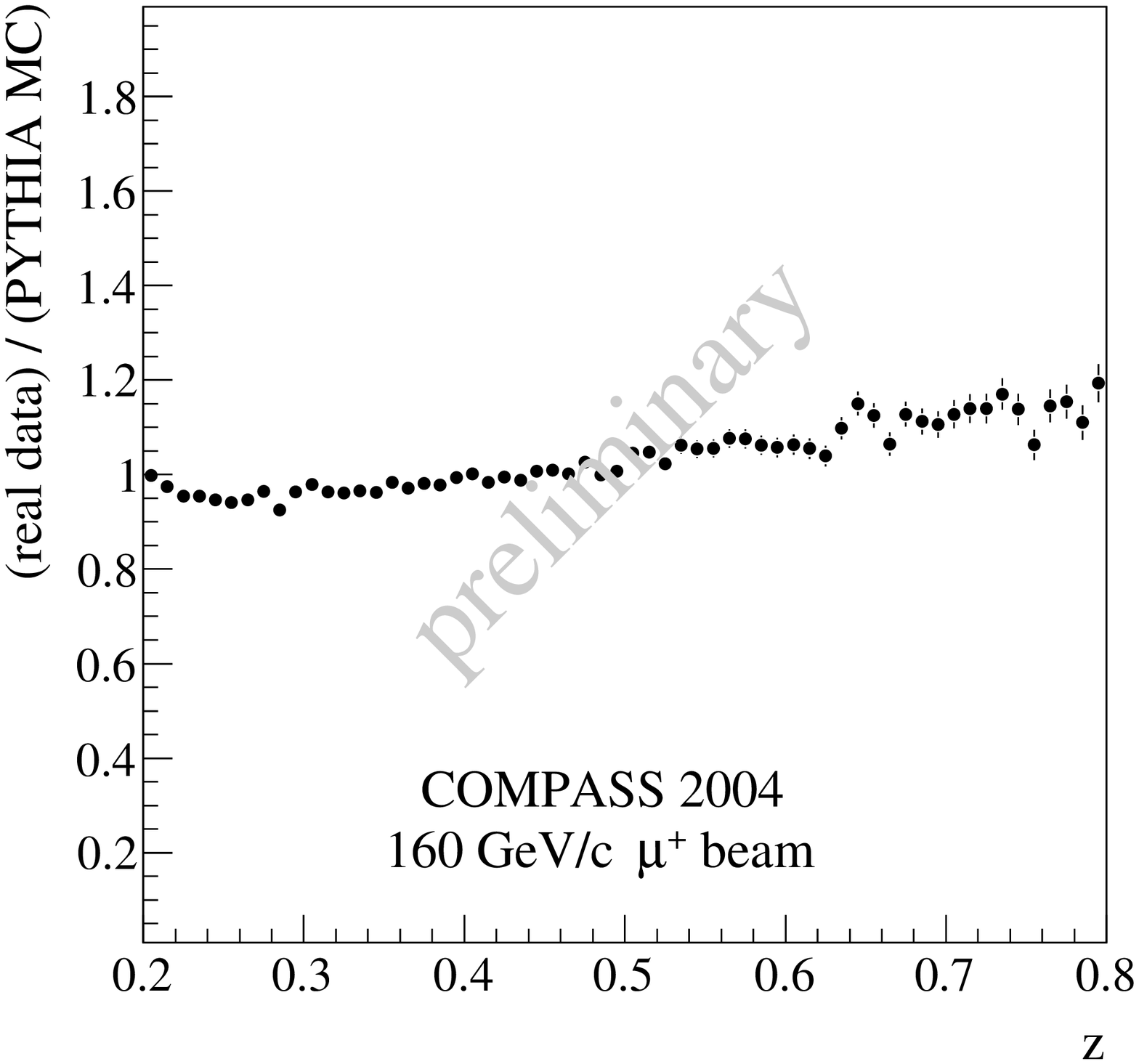}\includegraphics[width=14pc]{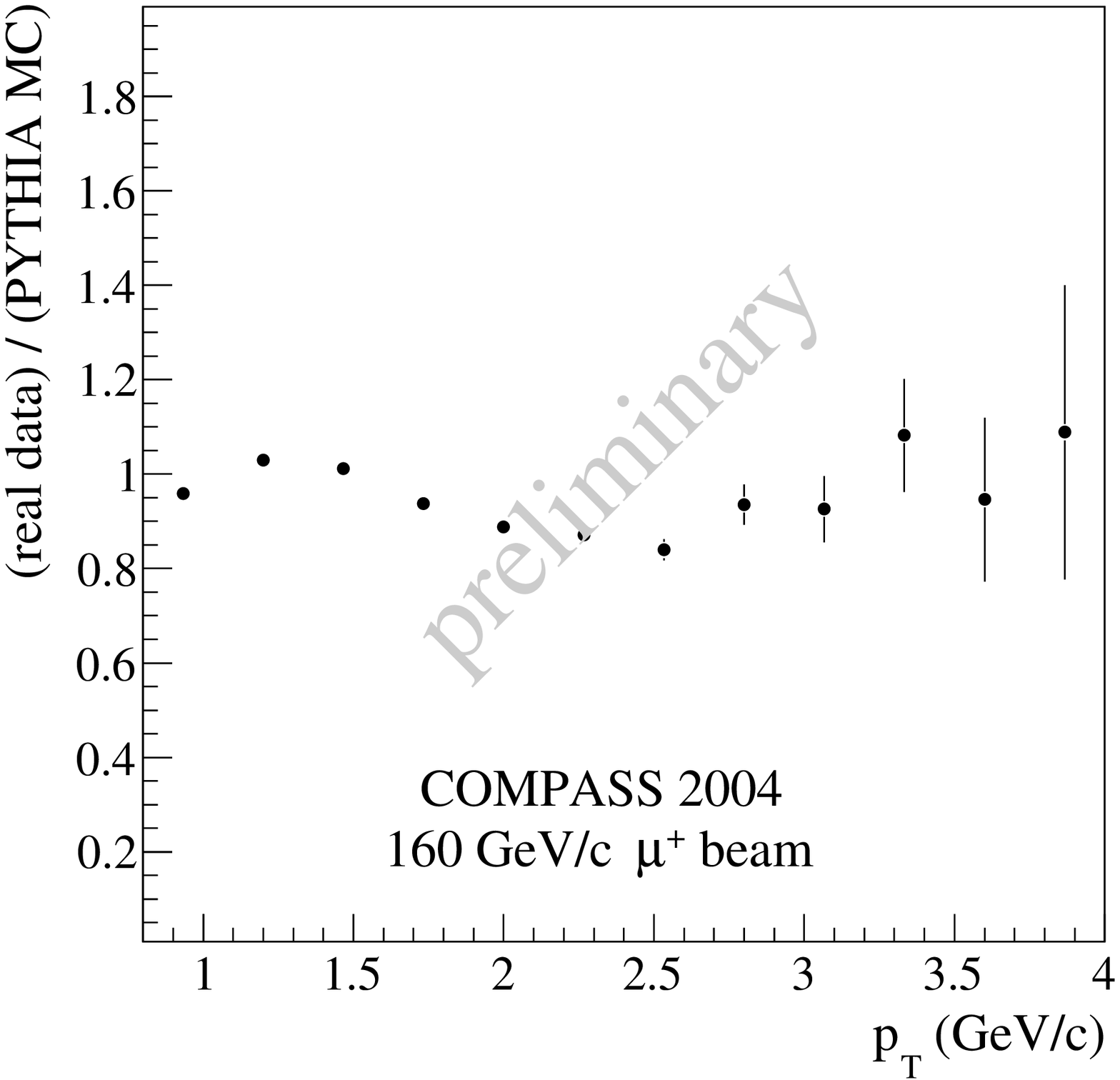}
\caption{\label{RealMC}. Ratio of real data under interest for the measurement over corresponding reconstructed simulated data. The figures corresponds to the description of $Q^2$ (top left), $y$ (top right), $z$ (bottom left) and $p_{T}$ (bottom right). Kinematic limits imposed include those described in~\ref{kine}}
\end{minipage} 
\end{figure}

A full simulation using the Monte-Carlo generator PYTHIA~\cite{pythia} was employed. The simulation generated $\mu^{+}$ proton and $\mu^{+}$ neutron interactions at appropriate center of mass energies. A GEANT~\cite{geant} based program which incorporates COMPASS detector materials and interactions was used. The GEANT based program simulated the response of the detector setup for the events generated by PYTHIA. The validity of the Monte-Carlo's description of the data was evaluated by estimating the data over Monte-Carlo ratio
in different kinematic and laboratory variables pertinent to the measurement. To accomplish this, histograms normalized by number of entries of data and Monte-Carlo were compared. The generated distributions described well the shapes of the real distributions as it can be inspected from Fig.~\ref{RealMC}. A good description of the experimental distribution in any given variable allows one to rely on the MC to extract the detector's acceptance and efficiency. This acceptance extraction can be done as a function the described variable in one dimension (1D). If the simulation can only poorly describe reality, a multi-dimensional ($>$2D) extraction of the acceptance is needed. Extraction of acceptances in both 1D as a function of $p_{T}$ and 2D  were performed and results were found compatible.

\subsection{Kinematic Distributions}\label{kine}
\begin{figure}[h]
\centering
\begin{minipage}{28pc}
\includegraphics[width=14pc]{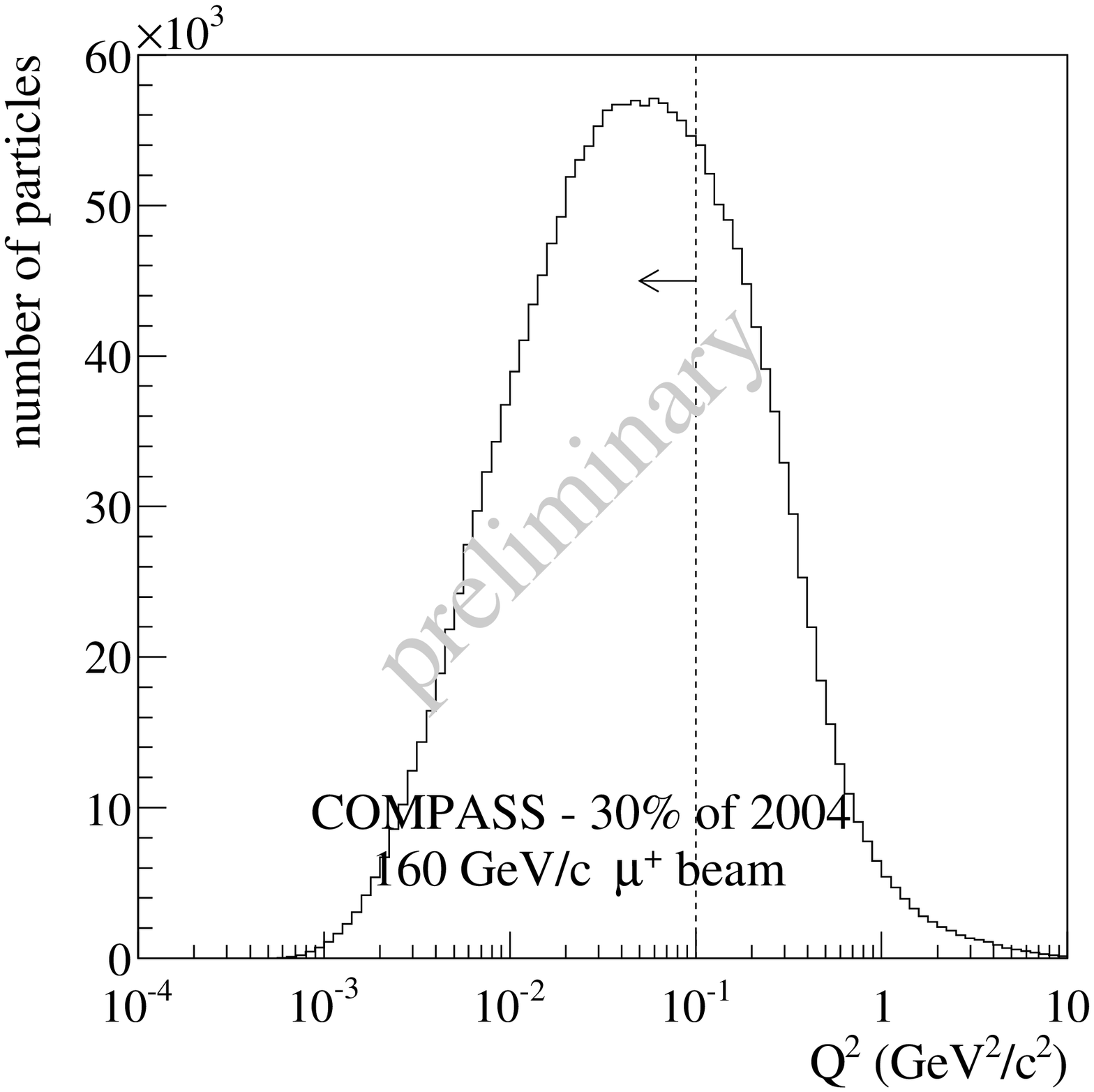}\includegraphics[width=14pc]{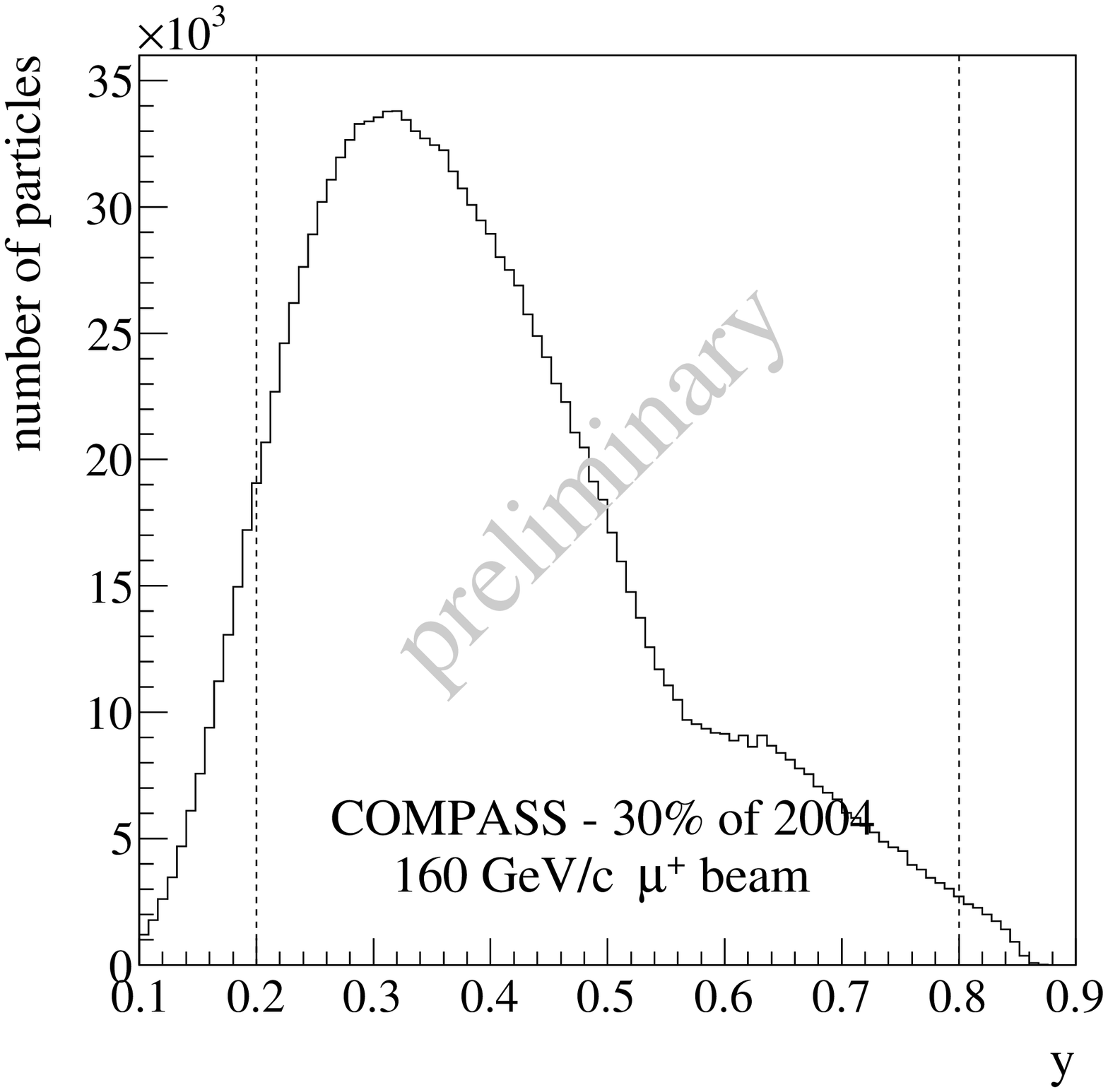}
\includegraphics[width=14pc]{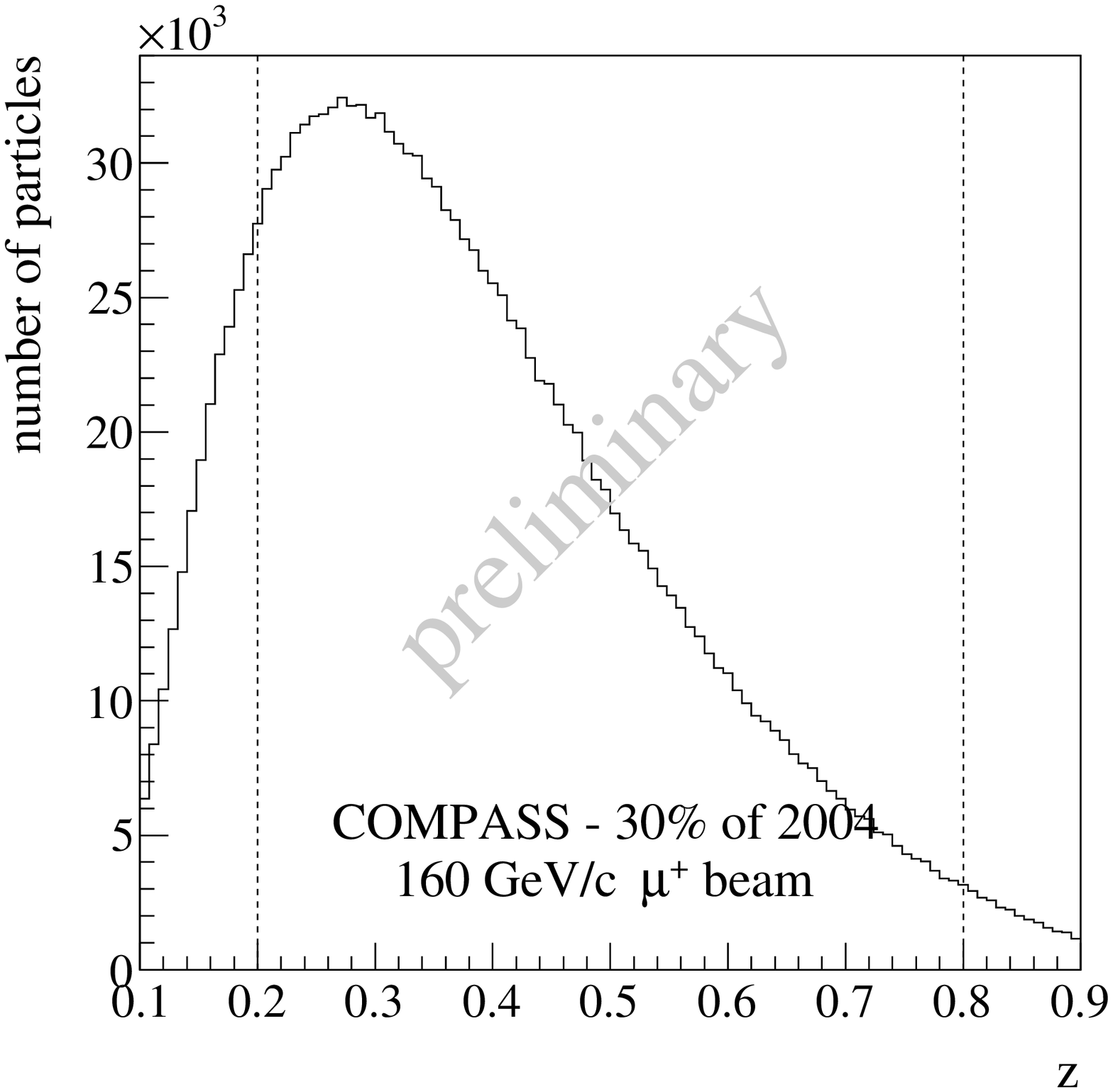}\includegraphics[width=14pc]{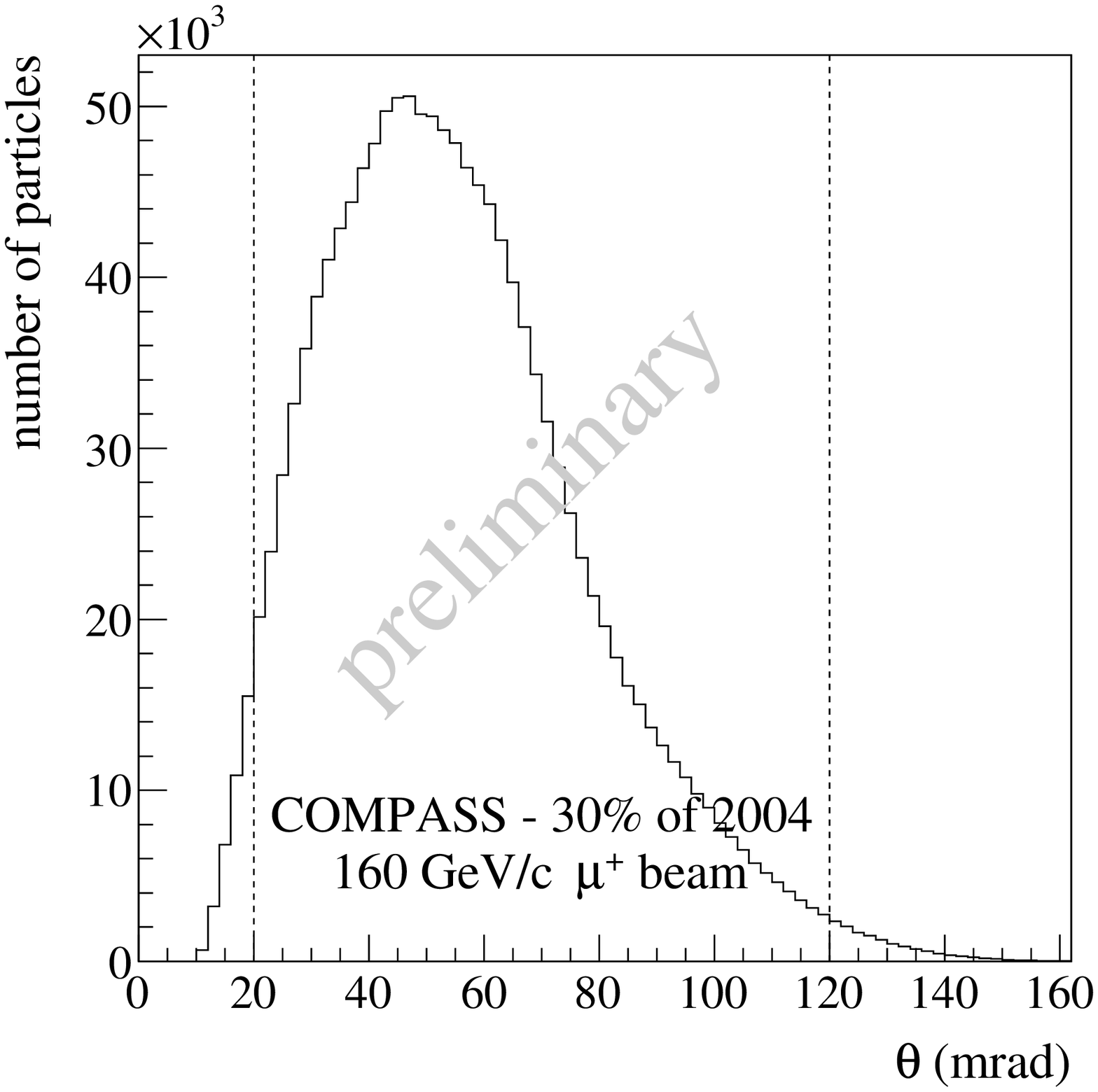}
\caption{\label{kinematics}Kinematic distributions of the 2004 data sample. The lines correspond to the kinematic limits imposed on the measurements.  The figures corresponds to the distribution of $Q^2$ (top left), $y$ (top right), $z$ (bottom left) and the scattering angle $\theta$(bottom right.)} 
\end{minipage} 
\end{figure}

The kinematic distributions of the measurements under interest can be found in figure~\ref{kinematics}. All pertinent detector and kinematic cuts are applied to the distributions shown in figure~\ref{kinematics},  except that for which the variable is being plotted. The lines indicate the kinematic limits imposed on the measurement. The main kinematic limits of the measurements are as follows:
\begin{itemize}
\item Incoming and scattered $\mu^{+}$ detected.
\item Energy transfer $y$ from $\mu^{+}$ to $\gamma$ between 0.2 and 0.8.
\item $Q^{2}$ Virtuality of $\gamma$ less than $0.1~GeV^{2}/c^{2}$.
\item Relative energy $z$ of produced hadron between 0.2 and 0.8.
\item Final state charged hadron detected with $p_{T}>\,1~GeV/c$
\item Scattering angle $\theta$ less than 120~mrad. 
\end{itemize}

\subsection{Integrated luminosity}
\begin{figure}[h]
\centering
\includegraphics[width=15pc]{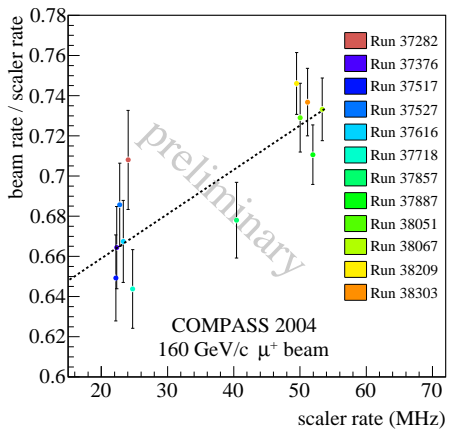}\hspace{2pc}%
\begin{minipage}[b]{14pc}\caption{\label{Lumi}Ratio of beam rates as measured with the random triggers and scalers. The observed dispersion of the ratio is due to dead time effects in the scaler measurements and its accounted for in the systematic uncertainty of the luminosity calculation.}
\end{minipage}
\end{figure}

The integrated luminosity is obtained from the beam flux measurement using only periods with stable
data taking and good spectrometer performance. This measurement was determined using beam scalers~\cite{compass} on a spill by spill basis, where each spill corresponds to the beam delivery by the SPS. The integrated luminosity  was found to be : $\int Ldt=$142.4 $pb^{-1}\,\pm 10\%$. The error is systematic and it reflects all uncertainties in the measurement including changes in beam intensity which affected the beam rate calculations (Fig.~\ref{Lumi}.)  

\subsection{Radiative corrections}
Radiative corrections for the production of hadrons at the kinematic limits described in section~\ref{kine} were evaluated~\cite{haprad} by Andrei Afanasev~\cite{andrei} from Jefferson Lab. His first estimates show that the data sample has a radiative effect contribution of less than $\sim 2\%$

\subsection{Hadron production at High $Q^{2}$}
\begin{figure}[h]
\begin{minipage}{16pc}
\includegraphics[width=15pc]{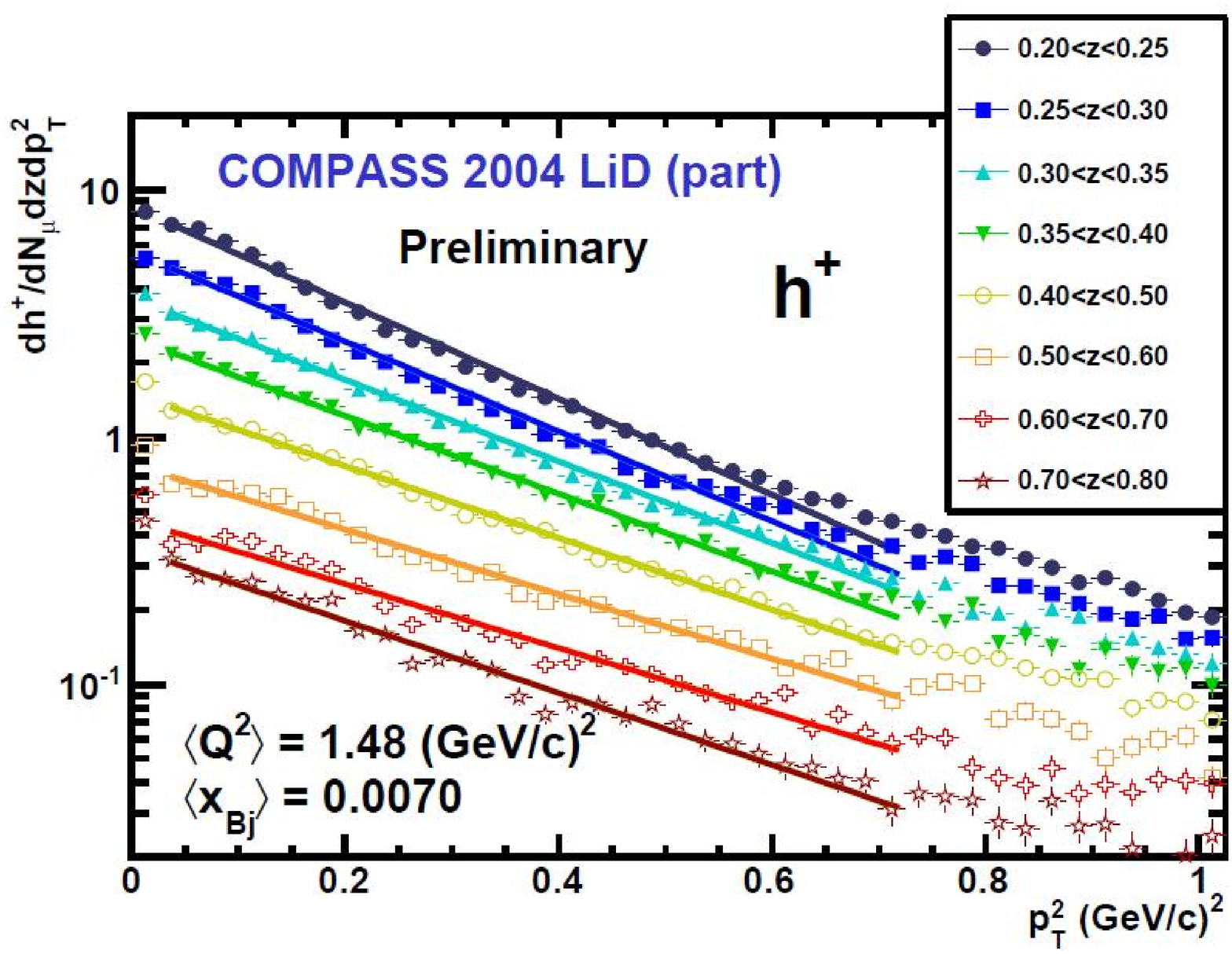}
\caption{\label{rajotte1}$p_{T}^{2}$ Spectra for $h^{+}$ at different values of  the relative hadron energy $z$. The spectra are fitted with a Gaussian function.}
\end{minipage}\hspace{2pc}%
\begin{minipage}{16pc}
\includegraphics[width=17pc]{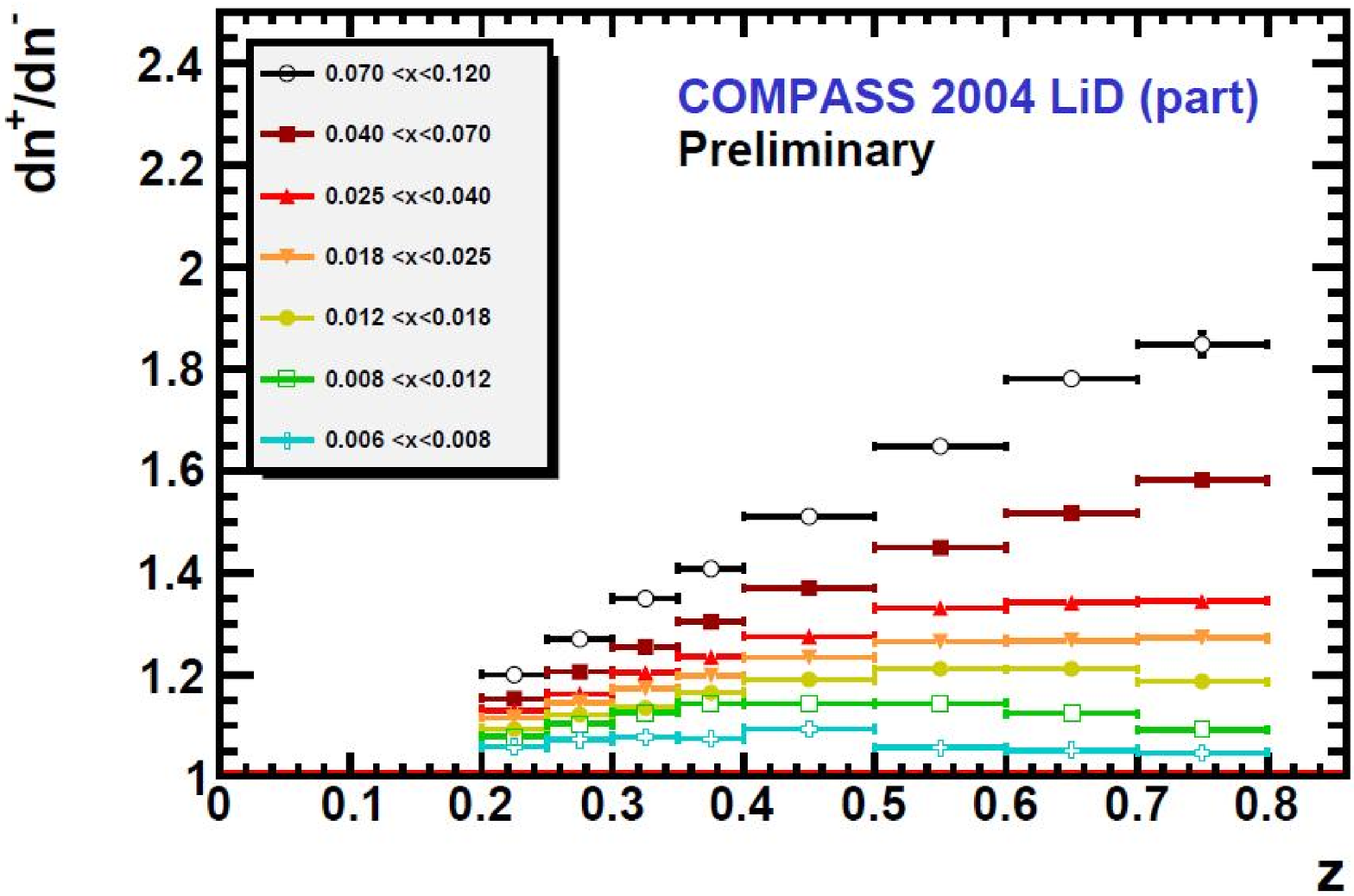}
\caption{\label{rajotte2}Ratio of $h^{+}$ to $h^{-}$ hadrons at $Q^{2}>1~GeV^{2}/c^{2}$ as a function of  $z$ at  different values of momentum fraction $x$.}
\end{minipage} 
\end{figure}

In addition to measuring cross-sections at low $Q^{2}$, COMPASS has released $p_{T}^{2}$ spectra measurements of charged hadrons at $Q^{2}>1~GeV^{2}/c^{2}$, $y$ between 0.1 and 0.9 and $z$ between 0.2 and 0.8.
These spectra (Fig.~\ref{rajotte1}) have have been fitted using a Gaussian function at low values of $p_{T}^2$ to determine the dependence of $<p_{T}^2> \,{\rm and\, } z$ among other kinematic variables. One primary goal of these measurements is to extract average intrinsic transverse momentum squared of partons
$<k_{T}^{2}>$ from fits of the $z$ dependence of the spectra as proposed in~\cite{mauro}. Similar to the case of the low $Q^2$ asymmetries, the unpolarized cross-section enters in the description of the weighted single spin asymmetry $A_{UT}^{sin(\phi\pi-\phi S)}$. In addition, the ratio of $h^{+}$ to  $h^{-}$ has also been measured as this value has sensitivity to the FF's. In the present measurements more $h^{+}$ are found in the measurements both at high and low $Q^{2}$ (Fig.~\ref{rajotte2}.)

\section{Conclusions and Remarks}
Measurements of unpolarized cross-sections at varying center of mass energies provide a benchmark test of the factorized sum-over-flavors NLO pQCD theory calculations. These unpolarized measurements in turn can provide validity for the framework currently used for the extraction of polarized gluon information from double spin asymmetries. Spectra at high $Q^2$ have been also measured as a function of several kinematical variables. These measurements aim at extracting intrinsic $<k_{T}^2>$ information and improve current transverse spin parametrizations.  At the time of the spin 2010 symposium, results for the measurements at low $Q^{2}$ had not yet released publicly.  Charged separated cross-sections results as well as the ratio of $h^{-}$ to $h^{+}$ have since been released and can be accessed through the COMPASS public page or by email request.

\subsection{Acknowledgments}
I thank Werner Vogelsang for his always useful discussions regarding particle production and for providing the updated NLO pQCD predictions. The collaborative work of Andrei Afanasev is also thanked as he provided the first radiative contributions calculations. 
This work was partially possible to a grant from the American National Science Foundation  in collaboration with the French Alternative Energies and Atomic Energy Commission.

\end{document}